\documentclass[12pt]{article}
\usepackage[utf8]{inputenc}
\usepackage[english]{babel}
\usepackage[centertags]{amsmath}
\usepackage{amssymb}
\usepackage{bm}
\usepackage{amsbsy}
\usepackage{graphicx}
\usepackage{appendix}
\usepackage{mathrsfs}
\usepackage{epsfig}
\usepackage{color}
\usepackage{euscript}
\usepackage{ulem}
\usepackage{multirow}
\usepackage{cite}

\definecolor{dgreen}{rgb}{0,0.6,0}

\definecolor{darkblue}{rgb}{0., 0, 1}

\definecolor{purple}{rgb}{0.65,0.,0.78}

\usepackage{jheppubm}
\newcommand{\nn}{\nonumber}
\newcommand{\be}{\begin{equation}}
\newcommand{\ee}{\end{equation}}
\newcommand{\bea}{\begin{eqnarray}}
\newcommand{\eea}{\end{eqnarray}}

\newcommand{\fg}{\mathfrak{g}}

\newcommand{\fB}{\mathfrak{B}}

\newcommand{\cL}{{\cal L}}
\newcommand{\cM}{{\cal M}}

\numberwithin{equation}{section}

\title{Einstein-dilaton-four-Maxwell Holographic Anisotropic Models}

\author{Irina Ya. Aref'eva$^a$, Kristina Rannu$^a$ and Pavel
  Slepov$^a$}

\affiliation{
  $^a$Steklov Mathematical Institute, Russian Academy of
  Sciences, \\ Gubkina str. 8, 119991, Moscow, Russia}
\emailAdd{arefeva@mi-ras.ru}
\emailAdd{rannu-ka@rudn.ru}
\emailAdd{slepov@mi-ras.ru}

\abstract{

In recent literature on holographic QCD the consideration of the five-dimensional Einstein-dilaton-Maxwell models has played a crucial role. Typically, one Maxwell field is associated with the chemical potential, while additional Maxwell fields are used to describe the anisotropy of the model. A more general scenario involves up to four Maxwell fields. The second field represents spatial longitudinal-transverse anisotropy, while the third and fourth fields describe anisotropy induced by an external magnetic field.
We consider an ansatz for the metric characterized by four functions at zero temperature and five functions at non-zero temperature. Maxwell field related to the chemical potential is treated with the electric ansatz, as is customary, whereas the remaining three Maxwell fields are treated with a magnetic ansatz.

We demonstrate that for the fully anisotropic diagonal metric only six out of seven equations are independent. One of the matter equations -- either the dilaton or the vector potential equation -- follows from the Einstein equations and the remaining matter equation. This redundancy arises due to the Bianchi identity for the Einstein tensor and the specific form of the stress-energy tensor in the model.
A procedure for solving this system of six equations is provided. This method generalizes previously studied cases involving up to three Maxwell fields. In the solution with three magnetic fields our analysis shows, that the dilaton equation is a consequence of the five Einstein equations and the equation for the vector potential.}

\keywords{AdS/QCD, holographic models, anisotropy, magnetic field}

\begin{document}

\maketitle

\section{Introduction}\label{introduction}

The holographic approach to studying QCD phase transitions and quark-gluon plasma (QGP) is currently being actively developed. Holography has emerged as a powerful tool for investigating ultrarelativistic heavy-ion collisions (HIC). Various experimental data such as transport coefficients, thermalization time, multiplicity, and direct-photon spectra can be effectively described within the framework of holographic QCD (HQCD). It is anticipated that holography will provide a more detailed structure of the QCD phase diagram. Different isotropic and anisotropic holographic models for describing QGP have been explored in
\cite{Casalderrey-Solana:2011dxg, Arefeva:2014kyw, DeWolfe:2013cua, Gursoy:2007cb, Gursoy:2007er, Gursoy:2008za, Gursoy:2010fj, Pirner:2009gr, He:2013qq, Arefeva:2014vjl, Yang:2015aia, Li:2017tdz, Arefeva:2018hyo, Chen:2018msc, Arefeva:2020vae, Bohra:2020qom, Arefeva:2020vhf, Arefeva:2020byn, Arefeva:2020bjk, Gursoy:2020kjd, Ballon-Bayona:2020xls, Li:2020hau, Arefeva:2021kku, Rannu:2021pcq, Arefeva:2021btm, Arefeva:2021mag, Dudal:2021jav, Caldeira:2021izy,  Arefeva:2021jpa, Hajilou:2021wmz, Arefeva:2022bhx, Shukla:2023pbp, Li:2016gfn, Gursoy:2016ofp, Dudal:2016joz, Gursoy:2017wzz, Gursoy:2018ydr, Bohra:2019ebj, Shahkarami:2021gzl, Arefeva:2022avn, Arefeva:2023jjh, Rannu:2024vrq}.\\

One class of the ``bottom-up'' HQCD models is based on the Einstein-dilaton-Maxwell action, where Maxwell field in the bulk supports chemical potential in the boundary field theory. The choice of the warp factor in the metric has a significant impact on the phase transition structure in HQCD. This choice should be guided by lattice results, such as those from the Columbia plot \cite{Brown:1990ev, Philipsen:2016hkv}.\\

Incorporating anisotropy into the holographic theory is essential as QGP is an anisotropic medium. It has been shown in \cite{Arefeva:2014vjl} that choosing a primary (spatial) anisotropy parameter value of approximately $\nu = 4.5$ accurately reproduces the energy dependence of the total multiplicity \cite{ALICE:2015juo}. Isotropic models, in contrast, cannot reproduce this behavior (see \cite{Arefeva:2014vjl} and references therein for more details). Additionally, understanding how primary anisotropy affects the QCD phase transition temperature is of great interest. Another form of anisotropy arises from the magnetic field, which also has significant effect on the QCD phase diagram. Experiments with relativistic HIC suggest that a very strong magnetic field, with $eB \sim 0.3$ GeV$^2$, is generated in the early stages of the collision \cite{Skokov:2009qp, Voronyuk:2011jd}. Therefore, investigating the effect of the magnetic field on the QCD features, particularly its phase transition diagram, is crucial for a deeper understanding of QCD.\\

We consider the Einstein-dilaton-four-Maxwell holographic models, which generalize the previously investigated holographic models with three Maxwell fields \cite{Arefeva:2020vae, Arefeva:2021jpa, Arefeva:2022bhx, Arefeva:2022avn, Arefeva:2023jjh, Rannu:2021pcq, Arefeva:2020bjk, Arefeva:2021btm, Arefeva:2021mag}.\\

This paper is organized as follows. 
In Sect.~\ref{section:model} the holographic model is presented. In Sect.~\ref{section:EE-2.15} and Sect.~\ref{section:FE-2.15} the Einstein equations and the matter field equations are considered in the parametrization \eqref{eq:2.12}.
In Sect.~\ref{section:BI} the consequences of the Bianchi identities are discussed, and it is shown that there are six independent equations of motion (EOM) in the model \eqref{eq:2.2}. In Sect.~\ref{section:sol} the scheme for finding a solution to this system of six equations is considered. 
In Sect.~\ref{section:concl} we summarize our main results. 
This work is complemented by the appendices. In Appendix~\ref{appA} the components of the Einstein tensor in the parametrization \eqref{eq:2.12} are presented. Appendix~\ref{appB} contains the expanded form of  the Einstein equations in the parametrization \eqref{eq:2.12}, while Appendix~\ref{appC} presents Ricci and Einstein tensors in the parametrization \eqref{eq:2.10}. Appendix~\ref{appD} provides the stress-energy tensor, and Appendix~\ref{appE} contains the EOM, all in the same parametrization \eqref{eq:2.10}.

\section{Model}
\label{section:model}

To construct an anisotropic holographic model we consider
the Einstein-dilaton-four-Maxwell system in the Einstein frame:
\bea 
 &&S = \int d^5x\, {\cL} =\int d^5x \left({\cL}_g - {\cL}_m \right) \label{eq:2.1} \\ 
 && = \int d^5x \, \sqrt{- g_5} \left[ R-
    \sum_{{\cM}=0}^{3} \cfrac{1}{4} \, f_{\cM}(\phi) F_{\cM}^2
    + \cfrac{1}{2} \, \partial_{\mu} \phi \, \partial^{\mu} \phi
    + V(\phi) \right], \label{eq:2.2}
\eea
where $F_{\cM}$ are Maxwell fields' stress tensors, ${\cM} = 0, 1, 2, 3$, $f_{\cM}(\phi)$ are the corresponding coupling functions, $\phi$ is the dilaton and $V(\phi)$ is its potential, $g_5$ is the determinant of the metric, and $R$
is its scalar curvature (Ricci scalar).

EOM for the metric, dilaton and Maxwell fields corresponding to the action \eqref{eq:2.2} have the form:
\begin{gather}
 G_{\mu\nu} = T_{\mu\nu}, \label{eq:2.3} \\
 D_\mu D^\mu \phi + V'(\phi) + \sum_{{\cM}=0}^{3} \cfrac{1}{4} \, f_{\cM}'(\phi) F_{\cM}^2 = 0, \label{eq:2.4} \\
  \partial_{\mu} (\sqrt{-g_5}\,f_{\cM}\, F_{\cM}^{\mu\nu}) = 0. \label{eq:2.5}
\end{gather}
where $G_{\mu\nu}$ is the Einstein tensor
\be
G_{\mu\nu} = R_{\mu\nu} - \cfrac12 \, g_{\mu\nu} R. \label{eq:2.6}
\ee
The full stress-energy tensor $T_{\mu\nu}$ consists of the dilaton and Maxwell parts:
\begin{gather}
  T_{\mu\nu} = T_{\mu\nu}^{\phi} + \sum_{{\cM}=0}^3 T_{\mu\nu}^{F_{\cM}}, \label{eq:2.7}
\end{gather}
where 
\begin{gather}
  T_{\mu\nu}^{\phi} = \partial_{\mu} \phi \, \partial_{\nu} \phi 
  - \cfrac{1}{2} \, g_{\mu\nu} g^{\sigma\kappa} 
  \partial_{\sigma} \phi \, \partial_{\kappa} \phi
  - g_{\mu\nu} V(\phi), \quad \label{eq:2.8} \\
  T_{\mu\nu}^{F_{\cM}} = \cfrac{1}{2} \,
  f_{\cM}(\phi) F_{\cM\,\mu\sigma} \, F_{\cM\,\nu}^{\ \sigma}
  - \cfrac{1}{4} \, g_{\mu\nu} \, f_{\cM}(\phi) (F_{\cM})^2. \label{eq:2.9}
\end{gather}
 
We search for the diagonal  solution:
\begin{gather}
  ds^2 = - \, g_{00} \, dt^2 + g_{11} \, dx_1^2
  + g_{22} \, dx_2^2 + g_{33} \, dx_3^2 + g_{44} \, dx_4^2, \label{eq:2.10}
\end{gather}
where metric components $g_{\mu\mu}$, $\mu = 0, 1, 2, 3, 4$, depend on the holographic coordinate $z$ only,
i.e.
\be
  g_{\mu\mu} = g_{\mu\mu}(z), \quad \mbox{where} \quad x_4 \equiv z. \label{eq:2.11}
\ee
We use also  the representation of the metric (\ref{eq:2.10}) in the following form:
\begin{gather}
  ds^2 = \fB^2(z) \left[
    - \, g(z) \, dt^2 
    + \fg_{1}(z) \, dx_1^2
    + \fg_{2}(z) \, dx_2^2
    + \fg_{3}(z) \, dx_3^2
    + \cfrac{dz^2}{g(z)} \right], \label{eq:2.12}
\end{gather}
where relations between $\fB(z)$, $g(z)$, $\fg_{i}(z)$, $i = 1, 2, 3$, and $g_{\mu\mu}$, $\mu = 0, 1, 2, 3, 4$, from (\ref{eq:2.10}) are the following:
\begin{gather}
  \begin{split}
    \fB^2(z) &= \sqrt{g_{00}(z) g_{44}(z)}, \quad 
    g = \sqrt{\cfrac{g_{00}(z)}{g_{44}(z)}}, \\
    \fg_1(z) = \cfrac{g_{11}(z)}{\sqrt{g_{00}(z) g_{44}(z)}}, &\quad
    \fg_2(z) = \cfrac{g_{22}(z)}{\sqrt{g_{00}(z) g_{44}(z)}}, \quad
    \fg_3(z) = \cfrac{g_{33}(z)}{\sqrt{g_{00}(z) g_{44}(z)}}. 
  \end{split}\label{eq:2.13}
\end{gather}
Metric (\ref{eq:2.12}) was used in works on anisotropic holographic QGP models \cite{Arefeva:2018hyo, Arefeva:2020vae, Arefeva:2020vhf, Arefeva:2020byn, Arefeva:2020bjk, Arefeva:2021kku, Rannu:2021pcq, Arefeva:2021btm, Arefeva:2021mag, Arefeva:2021jpa, Arefeva:2022bhx, Arefeva:2022avn, Arefeva:2023jjh, Rannu:2024vrq}.
\\

For the dilaton  we consider  the ansatz 
\begin{gather}
  \phi = \phi(z). \label{eq:2.14}
\end{gather}
For Maxwell fields we use either electric
\be\hspace{4pt}
  A_0 = A_t(z), \quad A_{i = 1,2,3,4} = 0, \hspace{103pt} \ \label{eq:2.15} 
\ee
or magnetic anzats
\be 
  F_1 = q_1 \, dx^2 \wedge dx^3, \ 
  F_2 = q_2 \, dx^1 \wedge dx^3, \
  F_3 = q_3 \, dx^1 \wedge dx^2, \label{eq:2.16}
\ee
where $M = 1, 2, 3$, $q_M$ are constant ``charges''
for Maxwell fields with magnetic anzats (\ref{eq:2.16}).
\\

\subsection{The Einstein Equations in Parametrization \eqref{eq:2.12}}\label{section:EE-2.15}

For the metric \eqref{eq:2.10}, or equivalently for \eqref{eq:2.12}, in accordance with the general theorem \cite{Dingle}, that for any diagonal metric depending on one coordinate only Ricci tensor is diagonal. Therefore, the Einstein tensor is diagonal as well.  
The explicit form of the Einstein tensor non-zero components $G_{\mu\nu}$ in parametrization \eqref{eq:2.12} are 
\bea
    G_{00} &=&  - \, \cfrac{g^2}{2} \left[
      \cfrac{6 \fB''}{\fB} 
      + \sum_{i=1}^3 \left(
        \cfrac{\fg_i''}{\fg_i} - \cfrac{\fg_i'^2}{2\fg_i^2} \right)
      + \cfrac{3 \fB'}{\fB} \,
      \cfrac{g'}{g} 
      +  \left( \cfrac{3\fB'}{\fB} + \cfrac{g'}{2g} \right)
      \sum_{i=1}^3 \cfrac{\fg_i'}{\fg_i}
      + \cfrac{1}{2} \, \sum_{i=1}^3 \prod_{j=1,j\ne i}^3 \cfrac{\fg_j'}{\fg_j} \right], \nn \\
 \label{eq:2.17} \\
 G_{ii} &=& \cfrac{g \, \fg_i}{2} \left[
      \cfrac{6 \fB''}{\fB} 
      + \cfrac{g''}{g}
      + \sum_{j=3,j\ne i}^1 \left(
        \cfrac{\fg_j''}{\fg_j} - \cfrac{\fg_j'^2}{2\fg_j^2} \right)
      + \cfrac{6 \fB'}{\fB} \,
      \cfrac{g'}{g}  + 
    \left( \cfrac{3\fB'}{\fB} + \cfrac{g'}{2g} \right)
      \sum_{j=1,j\ne i}^3 \cfrac{\fg_j'}{\fg_j}
      \right.\nn\\&+&\left. \cfrac{g'}{2g} \sum_{j=1,j\ne i}^3 \cfrac{\fg_j'}{\fg_j}
      + \cfrac{1}{2} \, \prod_{j=1,j\ne i}^3\cfrac{\fg_j'}{\fg_j} \right],
  \label{eq:2.18} \\
  G_{44} &=& \cfrac{6 \fB'^2}{\fB^2}
  + \cfrac{3 \fB'}{2 \fB} \, \cfrac{g'}{g}
  + \left( \cfrac{3\fB'}{\fB} + \cfrac{g'}{2g} \right)
  \sum_{i=1}^3 \cfrac{\fg_i'}{2\fg_i}
  + \sum_{i=1}^3 \prod_{j=1,j\ne i}^3 \cfrac{\fg_j'}{2 \fg_j}, \label{eq:2.19}
\eea
where $'= \partial/\partial z$ and $i = 1, 2, 3$.
The corresponding expressions for the parametrization \eqref{eq:2.10} are presented in Appendix~\ref{appC}.
\\

The non-zero components in the parametrization \eqref{eq:2.12} for the dilaton energy momentum tensor \eqref{eq:2.8} are
\begin{gather}
  \begin{split}
    T_{00}^{\phi}
      = \, &\cfrac{\fB^2 g}{2} \left(
      \cfrac{g \, \phi'^{\, 2}}{2 \fB^2} + V \right), \quad
  T_{44}^{\phi}
   = \cfrac{\fB^2}{2 g} \left(
      \cfrac{g \, \phi'^{\, 2}}{2 \fB^2} - V \right), \\
    &T_{ii}^{\phi}
    = - \, \cfrac{\fB^2 \fg_i}{2} \left(
      \cfrac{g \, \phi'^{\, 2}}{2 \fB^2} + V \right), \ 
    i = 1, 2, 3,
  \end{split}\label{eq:2.20}
\end{gather}
for the vector potential stress-energy tensor \eqref{eq:2.9} are
\begin{gather}
    T_{00}^{F_0}
      = \, \cfrac{g f_0 A_t'^2}{4 \fB^2}, \quad
    T_{ii}^{F_0}
    = \cfrac{\fg_i f_0 A_t'^2}{4 \fB^2}, \quad
  T_{44}^{F_0}
   = - \, \cfrac{f_0 A_t'^2}{4 \fB^2 g}, \ 
    i = 1, 2, 3,
    \label{eq:2.21}
\end{gather}
for the magnetic Maxwell fields stress-energy tensor \eqref{eq:2.9} are
\bea
    T_{00}^{F_i} 
    = \cfrac{g f_i q_i^2}{4 \fB^2 } \prod_{k=1,k \ne i}^3\frac{1}{\fg_k}, \quad
    T_{44}^{F_i} 
    = - \, \cfrac{f_i q_i^2}{4 \fB^2 g} \prod_{k=1,k \ne i}^3 \frac{1}{\fg_k}, \label{eq:2.22} \\
    T_{jj}^{F_i} 
    = (- 1)^{\delta_{ij}} \cfrac{\fg_j f_i q_i^2}{4 \fB^2} \prod_{k=1, k\ne i}^3\cfrac{1}{ \fg_k}, \ i, j= 1, 2, 3, \label{eq:2.23}
 \eea
where $(-1)^0 = 1$.
Therefore, our full system  of the Einstein equations contains only 5 equations:
\bea
    00: \,
    &-& \cfrac{12 \fB''}{\fB}
    - \cfrac{6 \fB'}{\fB} \, \cfrac{g'}{g}
    - \left( \cfrac{6 \fB'}{\fB} + \cfrac{g'}{g} \right) \sum_{i=1}^3 \cfrac{\fg_i'}{\fg_i}
    - \sum_{i=1}^3 \left(\cfrac{2 \fg_i''}{\fg_i} - \cfrac{\fg_i'^2}{\fg_i^2} \right)
    - \sum_{i=1}^3 \prod_{j=1,j\ne i} ^3\cfrac{\fg_j'}{\fg_j} \nn \\
    &=& \phi'^2 + \cfrac{2 \fB^2}{g} \, V 
    + \cfrac{f_0 A_t'^2}{\fB^2 g}
    + \cfrac{1}{\fB^2 g} \sum_{i=1}^3 f_i q_i^2\,\prod_{j=1,j\ne i}^3 \frac{1}{\fg_j},
   \label{eq:2.24} 
    \eea
    
    \bea
    ii: \,
    &-& \cfrac{12 \fB''}{\fB} 
    - \cfrac{2 g''}{g}
    - \cfrac{6 \fB'}{\fB} \left( \cfrac{2 g'}{g} + \sum_{j=1,j\ne i}^3 \cfrac{\fg_j'}{\fg_j} \right)
    - \sum_{j=1,j\ne i}^3 \left(\cfrac{2 \fg_j''}{\fg_j} - \cfrac{\fg_j'^2}{\fg_j^2} \right)
    - \left( \cfrac{2 g'}{g} \sum_{j=1,j\ne i}^3 \cfrac{\fg_j'}{\fg_j} + \prod_{j=1,j\ne i}^3 \cfrac{\fg_j'}{\fg_j} \right) \nn \\
    &=& \phi'^2 + \cfrac{2 \fB^2}{g} \, V 
    - \cfrac{f_0 A_t'^2}{\fB^2 g}
    + \cfrac{1}{\fB^2 g} \sum_{j=1}^3 (- 1)^{\delta_{ij}} f_j q_j^2\,\prod_{k=1,k\ne j}^3 \frac{1}{\fg_k}\,, \quad i  = 1, 2, 3, 
  \label{eq:2.25} 
      \eea 
      
    \bea
    44: \, 
    &-& \cfrac{24 \fB'^2}{\fB^2}
    - \cfrac{6 \fB'}{\fB} \, \cfrac{g'}{g}
    - \left( \cfrac{6 \fB'}{\fB} + \cfrac{g'}{g} \right) \sum_{i=1}^3 \cfrac{\fg_i'}{\fg_i}
    - \sum_{i=1}^3 \prod_{j=1,j\ne i}^3 \cfrac{\fg_j'}{\fg_j} \nn \\
    &=& - \, \phi'^2 + \cfrac{2 \fB^2}{g} \, V 
    + \cfrac{f_0 A_t'^2}{\fB^2 g}
    + \cfrac{1}{\fB^2 g} \sum_{i=1}^3 f_i q_i^2\,\prod_{j=1,j\ne i}^3 \frac{1}{\fg_j}.
   \label{eq:2.26}
\eea

\subsection{The Field Equations in Parametrization \eqref{eq:2.12}}\label{section:FE-2.15}

The EOM for the dilaton field (resulting from the variation of $\mathcal{L}_m$ with respect to the dilaton $\phi$) in the parametrization \eqref{eq:2.12} is:
\begin{gather}
  \begin{split}
 \cfrac{\delta {\cL}_m}{\delta \phi} \equiv \cfrac{\sqrt{-g_5 } \, g(z)}{\fB^2} &\left[
      \phi''
      + \phi' \left( \cfrac{3 \fB'}{\fB} 
        + \cfrac{g'}{g} 
        + \sum_{i=1}^3 \cfrac{\fg_i'}{2\fg_i} \right) 
        + \cfrac{\partial f_0}{\partial \phi} \, 
      \cfrac{A_t'^2}{2 \fB^2 g} \right. - \\ 
    &\quad \ \ - \sum_{i=1}^3 \cfrac{\partial f_i}{\partial \phi} \left. 
      \cfrac{q_i^2}{2 \fB^2 g} \prod_{j=1,j\ne i}^3 \frac{1}{\fg_j}
      - \cfrac{\fB^2}{g} \, \cfrac{\partial V}{\partial \phi} 
    \right] = 0. 
\end{split}\label{eq:2.27}
\end{gather}
The EOM for Maxwell fields with magnetic anzats (\ref{eq:2.16})
\begin{gather}
  \partial_\mu \left( 
    \sqrt{- \, g_5} \, f_M \, F_M^{\mu\nu}
  \right) = 0, \qquad M = 1, 2, 3, \label{eq:2.28}
\end{gather}
are satisfied automatically and do not contribute into our system, thus leaving us $7$ equations.

The EOM for the electric Maxwell $A_0$ 
(the result of the variation of $- \mathcal{L}_m$ with respect to the vector potential $A_t$) contains only one non-trivial component:

\begin{gather}
\cfrac{\delta {\cL}_m}{\delta A_t} \equiv \cfrac{\sqrt{-g_5} \, f_0}{\fB^4} \left[ A_t'' + A_t' \left( \cfrac{\fB'}{\fB} 
      + \cfrac{f_0'}{f_0}
      + \sum_{i=1}^3 \cfrac{\fg_i'}{2\fg_i} \right) \right] = 0.
  \label{eq:2.29}
\end{gather}

To summarize, for the ansatzes given in equations \eqref{eq:2.10}, \eqref{eq:2.14}--\eqref{eq:2.16} equations \eqref{eq:2.3}--\eqref{eq:2.5} yield 7 nontrivial relations.

In the next section we demonstrate that only 6 of these equations are independent. More precisely, if the Einstein equations are satisfied, then only one of the matter equations, for either the scalar or the electric field, is independent.

\section{Matter EOM and Bianchi Identities}\label{section:BI}

It is well known that the Einstein tensor satisfies the Bianchi identity \cite{TextBook},
\be
  \nabla_{\mu} G^{\mu\nu} = 0. \label{eq:3.01}
\ee
Thus, if the Einstein equation \eqref{eq:2.3} is satisfied, then it must also be the case that
\be
  \nabla_{\mu} T^{\mu\nu} = 0. \label{eq:3.02}
\ee
In other words, on the gravitational mass shell the stress-energy tensor satisfies the identity \eqref{eq:3.02}.
\\

\subsection{Covariant Derivative of the Dilaton Stress-Energy Tensor}

From the expression \eqref{eq:2.20}, taking into account \eqref{christoff}, we obtain the components of the covariant derivative of the dilaton stress-energy tensor as follows:
\begin{gather}
  \begin{split}
    &\nabla_0 T^{04}_{\phi}
    = \cfrac{g^2 \, \phi'^{\, 2}}{2 \fB^4} \left(
      \cfrac{\fB'}{\fB} + \cfrac{g'}{2 g} \right), \quad
    \nabla_i T^{i4}_{\phi} 
    = \cfrac{g^2 \, \phi'^{\, 2}}{2 \fB^4} \left(
      \cfrac{\fB'}{\fB} + \cfrac{\fg_i'}{2 \fg_i} \right), \\
    &\nabla_4 T^{44}_{\phi}
    = \cfrac{g^2 \, \phi'}{2 \fB^4} \left[
      \phi'' - \phi' \left(\cfrac{\fB'}{\fB} - \cfrac{g'}{2 g} \right)
      - \cfrac{\fB^2}{g} \, \cfrac{\partial V}{\partial \phi}
    \right], \ i, j = 1, 2, 3.
  \end{split}\label{eq:3.03} 
\end{gather}
All others components are zero 
\be
 \nabla_{\mu} T^{\mu\nu}_{\phi} \equiv 0, \quad \nu = 0, 1, 2, 3. \label{eomii} 
 \ee
These relations give us 
\begin{gather}
  \begin{split}
  \nabla_{\mu}T^{\mu\nu}_{\phi} 
    = \delta _4^\nu \, \nabla_{\mu}T^{\mu4}_{\phi}, \quad
    \nabla_{\mu}T^{\mu4}_{\phi} 
    = \cfrac{g^2 \phi'}{2 \fB^4} \left[
      \phi''
      + \phi' \left( \cfrac{3 \fB'}{\fB} 
        + \cfrac{g'}{g} 
        + \sum_{i=1}^3 \cfrac{\fg_i'}{2\fg_i} \right)
      - \cfrac{\fB^2}{g} \, \cfrac{\partial V}{\partial \phi}
    \right].
  \end{split}\label{eq:3.05}
\end{gather}

\subsection{Covariant Derivative of Maxwell Stress-Energy Tensor}

The covariant derivatives of Maxwell stress-energy tensors given by expressions \eqref{eq:2.21}--\eqref{eq:2.23} are
  \bea
&& \nabla_{\mu}T^{\mu4}_{F_0}
  = - \, \cfrac{g f_0 A_t'}{2 \fB^6} \left[
    A_t'' + A_t' \left( \cfrac{\fB'}{\fB} 
      + \cfrac{f_0'}{2 f_0}
      + \sum_{i=1}^3 \cfrac{\fg_i'}{2\fg_i} \right)
\right], \label{eq:3.06} \\
 &&   \nabla_{\mu}T^{\mu4}_{F_i} 
  = - \, \cfrac{g f_i' q_i^2}{4 \fB^6 }\,\prod_{j=1,j\ne i}^3 \frac{1}{\fg_j}, \quad i = 1, 2, 3. \label{eq:3.07}
 \eea
Hence, all Maxwell fields contributions to the covariant derivative of the stress-energy tensor summed up as  
 \bea
&&\nabla_{\mu}T^{\mu4}_{Maxwell}=
\nabla_{\mu}( T^{\mu\nu}_{F_0} + T^{\mu\nu}_{F_1} + T^{\mu\nu}_{F_2}
  + T^{\mu\nu}_{F_3}) \label{eq:3.08} \\
&&= - \, \cfrac{g }{2 \fB^6}\left \{  f_0 A_t' \left[
    A_t'' + A_t' \left( \cfrac{\fB'}{\fB} 
      + \cfrac{f_0'}{2 f_0}
      + \sum_{i=1}^3 \cfrac{\fg_i'}{2\fg_i} \right)
\right]+\sum _{i} \, \cfrac{f'_i q^2_i}{2} \prod_{j=1,j\ne i}^3 \frac{1}{\fg_j} \right\} \nn\\ 
&&= - \, \cfrac{g}{2 \fB^6}\left \{  f_0 A_t' \left[
    A_t'' + A_t' \left( \cfrac{\fB'}{\fB} 
      + \cfrac{f_0'}{f_0}
      + \sum_{i=1}^3 \cfrac{\fg_i'}{2\fg_i} \right)
\right]
+ \phi' \sum_{i} \frac{q_i^2\partial_\phi f_i}{2}\prod_{j=1,j\ne i}^3 \frac{1}{\fg_j} \right\}\nn
\\&&\quad + \, \cfrac{g (A'_t)^2 \phi' }{4 \fB^6} \, \frac{\partial f_0}{\partial \phi}. \nn
\eea
$\,$

\subsection{Covariant Derivative of the Total Matter Stress-Energy Tensor}

Using \eqref{eq:3.05} and \eqref{eq:3.08}
we get
\bea
&&\nabla_{\mu}T^{\mu4}_{total} = \nabla_{\mu}T^{\mu4}_{\phi}+\nabla_{\mu}T^{\mu4}_{Maxwell} \label{eq:3.09} \\
  &&= \cfrac{g^2 \phi'}{2 \fB^4} \left[
      \phi''
      + \phi' \left( \cfrac{3 \fB'}{\fB} 
        + \cfrac{g'}{g} 
        + \sum_{i=1}^3 \cfrac{\fg_i'}{2\fg_i} \right)
      - \cfrac{\fB^2}{g} \, \cfrac{\partial V}{\partial \phi}
    - \frac{1}{2g\fB^2}\sum _{i} q_i^2\partial_\phi f_i \prod_{j=1,j\ne i}^3 \frac{1}{\fg_j}
    + \cfrac{(A_t')^2}{2 g\fB^2} \, \frac{\partial f_0}{\partial \phi} \right] \nn\\
    &&- \cfrac{g f_0 A_t'}{2 \fB^6} \left[
    A_t'' + A_t' \left( \cfrac{\fB'}{\fB} 
      + \cfrac{f_0'}{f_0}
      + \sum_{i=1}^3 \cfrac{\fg_i'}{2\fg_i} \right)
\right]. \nn
\eea

Note that due to \eqref{eq:2.27} and \eqref{eq:2.29} the expressions on the RHS of \eqref{eq:3.09} are nothing but combinations of the RHS of the EOM for the dilaton and electric fields, multiplied by $\phi'$ and $A'_t$, respectively, with some overall factors, i.e.
\be  
   \nabla_{\mu} T_{total}^{\mu\nu}
    = \cfrac{g}{2 \sqrt{-g_5} \, \fB^2} \left(
      A_t' \,  \cfrac{\delta {\cL}}{\delta A_t} +
      \phi' \, \cfrac{\delta {\cL}}{\delta \phi} \right). \label{eq:3.10} 
\ee

Therefore, on a solution of the Einstein equation the LHS of \eqref{eq:3.10} must be zero, since the Bianchi identity holds for the Einstein tensor. If we then assume, that one of the matter field equations is satisfied, the validity of the second equation follows from the vanishing of the covariant derivative of the total stress-energy tensor. 
Altogether, we have seven equations: five Einstein equations and two matter field equations. However, from the considerations above it follows, that there are only six independent EOM: five Einstein equations \eqref{eq:2.24}–\eqref{eq:2.26} and a matter field equation.
Usually, the electric potential equation is considered as the independent one.
      
\section{Remarks on EOM Solution for the Model \eqref{eq:2.2}}\label{section:sol}

If Maxwell coupling functions and the scalar potential are given, one has 6 equations for 7 functions: 5 metric functions, the dilaton $\phi$, and the electric potential $A_t$. 
We can fix one function (in the parametrization \eqref{eq:2.10}) to be 1. This can be achieved by rescaling $z$: $g_{44}dz^2 \to d\bar{z}^2$. Therefore, we need to define 4 metric functions and 2 matter functions: $\phi$ and $A_t$ from the set of 6 EOM.
\\

Another approach can also be employed: one can specify the metric functions $\fB^2$ and $\fg_i(z)$, $i = 1, 2, 3$, and search for two Maxwell coupling functions $f_{\cM}$, with ${\cM} = 2, 3$, the blackening function $g$, the electric potential $A_t$, the scalar potential $V(\phi)$, and the dilaton $\phi$ itself. This approach assumes that two of Maxwell potentials are given, in this case $f_0$ and $f_1$.
\\

In holographic applications the general model with four Maxwell fields has not yet been explored. The more complicated model used so far involves three Maxwell fields \cite{Arefeva:2022avn, Arefeva:2020vae}, which corresponds to setting $q_2 = 0$ and assuming $\fg_{1} = 1$. In this case we used the second approach to construct solutions of the EOM. We specified the form of the prefactor $\fB$ and the function $f_0$ (with $f_2 = 0$). The system was set up by the Einstein equation and the electric field equation, while the dilaton equation was treated as a constraint. The form of $f_0$ was chosen to reproduce the meson Regge spectrum \cite{He:2013qq, Yang:2015aia, Li:2017tdz}. \\

To solve the EOM system, it is convenient to combine the Einstein equations \eqref{ein00}-\eqref{ein44} into linear combinations, thus excluding the repeating terms. To achieve this, we use the procedure that consists of writing the combination of EOM \eqref{linear}, indicated in Appendix~\ref{app:UFEOM}. We get
\bea
 g'' &+& g' \left( \cfrac{3 \fB'}{\fB} - \cfrac{\fg_1'}{2 \fg_1} + \cfrac{\fg_2'}{2 \fg_2} + \cfrac{\fg_3'}{2 \fg_3} \right)
    - g \left[ 
    \cfrac{\fg_1''}{\fg_1} - \cfrac{\fg_1'^2}{2 \fg_1} + \cfrac{\fg_1'}{\fg_1} \left( 
    \cfrac{3 \fB'}{\fB} + \cfrac{\fg_2'}{2 \fg_2} + \cfrac{\fg_3'}{2 \fg_3} \right) \right] \nn \\
    &-& \cfrac{f_0 A_t'^2}{\fB^2} 
    - \cfrac{f_2 q_2^2}{\fB^2 \fg_1 \fg_3}
    - \cfrac{f_3 q_3^2}{\fB^2 \fg_1 \fg_2} = 0,
   \label{eq:4.01} 
   \\\,\nn\\
     \cfrac{6 \fB''}{\fB} 
    &-& \cfrac{12 \fB'^2}{\fB^2}
    + \sum_{i=1}^3 \left( 
    \cfrac{\fg_i''}{\fg_i} - \cfrac{\fg_i'^2}{\fg_i^2}
    \right)
    + \phi'^2 = 0, \label{eq:4.02} \\
    \cfrac{\fg_1''}{\fg_1} &-& \cfrac{\fg_1'^2}{2 \fg_1^2}
    - \left( \cfrac{\fg_2''}{\fg_2} - \cfrac{\fg_2'^2}{2 \fg_2^2} \right)
    + \left( \cfrac{\fg_1'}{\fg_1} - \cfrac{\fg_2'}{\fg_2} \right)
    \left( \cfrac{3 \fB'}{\fB} + \cfrac{g'}{g} + \cfrac{\fg_3'}{2 \fg_3} \right)
    - \cfrac{f_1 q_1^2}{\fB^2 g \fg_2 \fg_3}
    + \cfrac{f_2 q_2^2}{\fB^2 g \fg_1 \fg_3} = 0, \nn\\ \label{eq:4.03} \\
    \cfrac{\fg_1''}{\fg_1} &-& \cfrac{\fg_1'^2}{2 \fg_1^2}
    - \left( \cfrac{\fg_3''}{\fg_3} - \cfrac{\fg_3'^2}{2 \fg_3^2} \right)
    + \left( \cfrac{\fg_1'}{\fg_1} - \cfrac{\fg_3'}{\fg_3} \right)
    \left( \cfrac{3 \fB'}{\fB} + \cfrac{g'}{g} + \cfrac{\fg_2'}{2 \fg_2} \right)
    - \cfrac{f_1 q_1^2}{\fB^2 g \fg_2 \fg_3}
    + \cfrac{f_3 q_3^2}{\fB^2 g \fg_1 \fg_2} = 0, \nn\\ \label{eq:4.04} \\
   \cfrac{6 \fB''}{\fB} 
    &+& \cfrac{12 \fB'^2}{\fB^2}
    + \cfrac{g''}{g}
    + \cfrac{\fg_1''}{\fg_1} - \cfrac{\fg_1'^2}{2 \fg_1^2}
    + \cfrac{\fg_3''}{\fg_3} - \cfrac{\fg_3'^2}{2 \fg_3^2}
    + \cfrac{3 \fB'}{\fB} \left( \cfrac{3 g'}{g} - \cfrac{\fg_2'}{\fg_2} \right)
    - \cfrac{g'}{g} \, \cfrac{\fg_2'}{\fg_2}
    + \cfrac{\fg_1'}{2 \fg_1} \, \cfrac{\fg_3'}{\fg_3} \, \nn \\
    &+& 3 \left( \cfrac{2 \fB'}{\fB} + \cfrac{g'}{2 g} \right) \sum_{i=1}^3 \cfrac{\fg_i'}{\fg_i}
    + \cfrac{1}{2} \sum_{i=1}^3 \prod_{j=1,j\ne i}^3 \cfrac{\fg_j'}{\fg_j} 
    + \cfrac{f_2 q_2^2}{\fB^2 g \fg_1 \fg_3}
    + \cfrac{2 \fB^2}{g} \, V = 0.
    \label{eq:4.05}
\eea

Assuming the function $f_1$ to be given, we can find $f_2$ and $f_3$ from (\ref{eq:4.03}) and (\ref{eq:4.04}), respectively. Substituting the obtained expressions into equation \eqref{eq:4.01}, we get the equation for the blackening function $g$ in the form
\bea
    g'' &+& g' \left( \cfrac{3 \fB'}{\fB} + \cfrac{3 \fg_1'}{2 \fg_1} - \cfrac{\fg_2'}{2 \fg_2} - \cfrac{\fg_3'}{2 \fg_3} \right) \label{eq:4.06} \\
    &-& g \left[ 
    - \, \cfrac{\fg_1''}{\fg_1} + \cfrac{\fg_1'^2}{2 \fg_1}
    + \cfrac{\fg_2''}{\fg_2} - \cfrac{\fg_2'^2}{2 \fg_2}
    + \cfrac{\fg_3''}{\fg_3} - \cfrac{\fg_3'^2}{2 \fg_3}
    + \cfrac{3 \fB'}{\fB} \left( - \, \cfrac{\fg_1'}{\fg_1} + \cfrac{\fg_2'}{\fg_2} + \cfrac{\fg_3'}{\fg_3} \right)
    + \cfrac{\fg_2'}{\fg_2} \, \cfrac{\fg_3'}{\fg_3}
    \right] \nn\\
    &-& \cfrac{f_0 A_t'^2}{\fB^2} 
    - \cfrac{2 f_1 q_1^2}{\fB^2 \fg_2 \fg_3} = 0, \nn
\eea
where $f_1$ turns out to be arbitrary and needs to be fixed, while $A_t'$ is determined from the electric field equation \eqref{eq:2.29}. Analogously, any two magnetic coupling functions can be expressed in terms of the third one, so either $f_2$ or $f_3$ can also be chosen arbitrarily. The choice of the arbitrary coupling function is determined by physical considerations and depends on the specific problem investigated.
\\

After the blackening function is found, we can determine the potential as a function of $z$ from equation \eqref{eq:4.05}.
From equation \eqref{eq:4.02} we know $\phi'$ and therefore $\phi = \phi(z)$. If the obtained function is monotonic, we can recover $V = V(\phi)$. The same strategy has been applied in the cases of one, two, and three Maxwell fields in \cite{Yang:2015aia,Li:2017tdz, Arefeva:2018hyo}, and \cite{Arefeva:2022avn,Arefeva:2020vae}, respectively.

Note, that similar method can be also used for more complicated scalar field models \cite{Ahmed:2023zkk, Ahmed:2024rbj, Liu:2023pbt}.

\section{Conclusions}\label{section:concl}

We have considered the Einstein-dilaton-four-Maxwell holographic models in 5-dimen\-sional space-time. It has been shown that, in the fully anisotropic case, there are only six independent equations out of seven due to the Bianchi identities for the Einstein tensor. A scheme for finding a solution to the system of six equations, which generalizes previously considered cases involving up to three Maxwell fields, has been presented.

\section*{Acknowledgments}

This work is supported by Russian Science Foundation grant
20-12-00200.

\newpage

\appendix 

\section*{Appendix}

\section{Ricci and Einstein Tensors in Parametrization \eqref{eq:2.12}}\label{appendixA}
\label{appA}

In parametrization (\ref{eq:2.12}) non-zero Christoffel components are 
\begin{gather}
  \begin{split}
    \Gamma^0_{40} &= \cfrac{\fB'(z)}{\fB(z)} 
    + \cfrac{g'(z)}{2 g(z)}, \qquad
    \Gamma^i_{4i} = \cfrac{\fB'(z)}{\fB(z)} 
    + \cfrac{\fg_i(z)}{2 \fg_i(z)}, \ i = 1,2,3, \\ 
    \Gamma^4_{00} &= g(z)^2 \left( \cfrac{\fB'(z)}{\fB(z)} +
      \cfrac{g'(z)}{2 g(z)} \right), \qquad
    \Gamma^4_{44} = \cfrac{\fB'(z)}{\fB(z)} 
    - \cfrac{g'(z)}{2 g(z)}, \\
    &\Gamma^4_{ii} = - \, g(z) \fg_i(z) \left( \cfrac{\fB'(z)}{\fB(z)}
      + \cfrac{\fg_i(z)}{2 \fg_i(z)} \right), \ i =
    1,2,3.
  \end{split}\label{christoff}
\end{gather}
Ricci tensor and Ricci scalar are
\begin{gather}
  R_{00} = \cfrac{g}{2} \left[
    g'' + g' \left( 
      \cfrac{5 \fB'}{\fB}
      + \sum_{i=1}^3 \cfrac{\fg_i'}{2\fg_i} \right)
    + 2 g \left( \cfrac{\fB''}{\fB}
      + \cfrac{2 \fB'^2}{\fB^2} 
        + \cfrac{\fB'}{\fB} \sum_{i=1}^3 \cfrac{\fg_i'}{2\fg_i} \right)
  \right], \label{ricci00} \\
  R_{ii} = - \, \cfrac{g}{2} \left[
    \fg_i'' + \fg_i' \left( 
      \cfrac{5 \fB'}{\fB}
      + \cfrac{g'}{g}
      - \cfrac{\fg_i'}{2\fg_i}
      + \sum_{j\ne i} \cfrac{g_j'}{2g_j} \right)
    + 2 \fg_i \left\{ \cfrac{\fB''}{\fB}
      + \cfrac{2 \fB'^2}{\fB^2}
        + \cfrac{\fB'}{\fB} \left( \cfrac{g'}{g}
        - \cfrac{\fg_i'}{2\fg_i}
        + \sum_{j\ne i} \cfrac{g_j'}{2g_j} \right) \right\}
  \right], \nn\\ i = 1,2,3 \label{ricciii} \\
  R_{44} = - \, \cfrac{1}{2} \left[
    \cfrac{g''}{g} + \sum_{i=1}^3 \left( \cfrac{\fg_i''}{\fg_i} 
      - \cfrac{\fg_i'^2}{2\fg_i^2} \right)
    + \cfrac{8\fB''}{\fB}
    - \cfrac{8\fB'^2}{\fB^2}
    + \cfrac{5\fB'}{\fB} \, \cfrac{g'}{g}
    + \left( \cfrac{\fB'}{\fB} + \cfrac{g'}{2g} \right)
    \sum_{i=1}^3 \cfrac{\fg_i'}{\fg_i} \right], \label{ricci44} \\
   R = - \, \cfrac{g}{\fB^2} \left[
      \cfrac{g''}{g} + \sum_{i=1}^3 \left(
        \cfrac{\fg_i''}{\fg_i} - \cfrac{\fg_i'^2}{2\fg_i^2} \right)
      + \cfrac{8\fB''}{\fB}
      + \cfrac{4\fB'^2}{\fB^2}
      + \cfrac{8\fB'}{\fB} \, \cfrac{g'}{g} + 
    \left( \cfrac{4\fB'}{\fB} + \cfrac{g'}{g} \right) 
      \sum_{i=1}^3 \cfrac{\fg_i'}{\fg_i}
      + \cfrac{1}{2} \,
      \sum_{i=1}^3 \prod_{j\ne i} \cfrac{\fg_j'}{\fg_j}
      \right].
  \label{riccisc}
\end{gather}

\section{Expanded Form of Einstein Equations \eqref{eq:2.24}--\eqref{eq:2.26}}
\label{appB}

\bea
 \mbox{\bf 00:}&\,&   g^2   \phi'^2+2 \fB^2 g   V+ \frac{ g f_0 A_t'^2}{\fB^2}  + \frac{g f_1 q_1^2 }{\fB^2 \fg_2 \fg_3}
+ \cfrac{g f_2 q_2^2}{\fB^2 \fg_1 \fg_3}
    + \cfrac{g f_3 q_3^2}{\fB^2 \fg_1 \fg_2} \label{00-math} \\  \nn\\
    &&=  - \, \frac{\fB'' g^2}{\fB} - \cfrac{6 \fB' g' g}{\fB} -\frac{6 \fB' g^2}{\fB} \Big( \frac{\fg_1'}{\fg_1} + 
    \frac{\fg_2'}{\fg_2}+\frac{\fg_3'}{\fg_3}\Big)
    - g' g \Big(\frac{\fg_1'}{\fg_1}
   +\frac{ \fg_2' }{\fg_2} +
   \frac{ \fg_3' }{\fg_3}\Big) \nn \\\nn\\
    &&\quad - \, 2g^2\Big(
  \frac{\fg_1''}{\fg_1}+\frac{\fg_2''}{\fg_2}+\frac{\fg_3''}{\fg_3}\Big) 
  + g^2\Big(
  \frac{\fg_1'^2}{\fg_1^2}+\frac{\fg_2'^2}{\fg_2^2}+\frac{\fg_3'^2}{\fg_3^2}\Big)
  - g^2\Big(\frac{\fg_1' \fg_2' }{\fg_1 \fg_2}+\frac{\fg_1' \fg_3' }{\fg_1 \fg_3}+\frac{\fg_2' \fg_3' }{\fg_2 \fg_3}\Big), \nn
\eea
\bea
 \mbox{\bf 11:}&\,& g \fg_1 \phi'^2+2 \fB^2 \fg_1   V
 - \frac{ \fg_1 f_0 A_t'^2}{\fB^2}  
 +  \frac{\fg_1 f_1 q_1^2 }{\fB^2 \fg_2 \fg_3}
- \cfrac{f_2 q_2^2}{\fB^2  \fg_3}
    - \cfrac{f_3 q_3^2}{\fB^2 \fg_2} \label{11-math} \\  \nn\\
    &&=  - \, \frac{12  
    \fB'' g \fg_1}{\fB} 
    - 2 g'' \fg_1
    - \cfrac{12 \fB' g' \fg_1}{\fB} -\frac{6 \fB' g \fg_1}{\fB} \Big( 
    \frac{\fg_2'}{\fg_2}+\frac{\fg_3'}{\fg_3}\Big)
    - 2 g' \fg_1 \Big(\frac{ \fg_2' }{\fg_2} +
   \frac{ \fg_3' }{\fg_3}\Big) \nn \\\nn\\
    &&\quad - \, 2 g \fg_1 \Big(
  \frac{\fg_2''}{\fg_2}+\frac{\fg_3''}{\fg_3}\Big) 
+ g \fg_1 \Big(
  \frac{\fg_2'^2}{\fg_2^2}+\frac{\fg_3'^2}{\fg_3^2}\Big)
  - g \fg_1 \, \frac{\fg_2' \fg_3' }{\fg_2 \fg_3}, \nn \\
  \mbox{\bf 22:}&\,& g \fg_2 \phi'^2+2 \fB^2 \fg_2   V
 - \frac{ \fg_2 f_0 A_t'^2}{\fB^2}  
 -  \frac{f_1 q_1^2 }{\fB^2 \fg_3}
+ \cfrac{\fg_2 f_2 q_2^2}{\fB^2 \fg_1 \fg_3}
    - \cfrac{f_3 q_3^2}{\fB^2 \fg_1} \label{22-math} \\  \nn\\
    &&=  - \, \frac{12  
    \fB'' g \fg_2}{\fB} 
    - 2 g'' \fg_2
    - \cfrac{12 \fB' g' \fg_2}{\fB} -\frac{6 \fB' g \fg_2}{\fB} \Big( 
    \frac{\fg_1'}{\fg_1}+\frac{\fg_3'}{\fg_3}\Big)
    - 2 g' \fg_2 \Big(\frac{ \fg_1' }{\fg_1} +
   \frac{ \fg_3' }{\fg_3}\Big) \nn \\\nn\\
    &&\quad - \, 2 g \fg_2 \Big(
  \frac{\fg_1''}{\fg_1}+\frac{\fg_3''}{\fg_3}\Big) 
   + g \fg_2 \Big(
  \frac{\fg_1'^2}{\fg_1^2}+\frac{\fg_3'^2}{\fg_3^2}\Big)
  - g \fg_2 \, \frac{\fg_1' \fg_3' }{\fg_1 \fg_3}, \nn \\
 \mbox{\bf 33:}&\,& g \fg_3 \phi'^2+2 \fB^2 \fg_3   V
 - \frac{ \fg_3 f_0 A_t'^2}{\fB^2}  
 -  \frac{f_1 q_1^2 }{\fB^2 \fg_2}
- \cfrac{f_2 q_2^2}{\fB^2 \fg_1}
    + \cfrac{\fg_3 f_3 q_3^2}{\fB^2 \fg_1 \fg_2} \label{33-math} \\  \nn\\
    &&=  - \, \frac{12  
    \fB'' g \fg_3}{\fB} 
    - 2 g'' \fg_3
    - \cfrac{12 \fB' g' \fg_3}{\fB} -\frac{6 \fB' g \fg_3}{\fB} \Big( 
    \frac{\fg_1'}{\fg_1}+\frac{\fg_2'}{\fg_2}\Big)
    - 2 g' \fg_3 \Big(\frac{ \fg_1' }{\fg_1} +
   \frac{ \fg_2' }{\fg_2}\Big) \nn \\\nn\\
    &&\quad - \, 2 g \fg_3 \Big(
  \frac{\fg_1''}{\fg_1}+\frac{\fg_2''}{\fg_2}\Big) 
  + g \fg_3 \Big(
  \frac{\fg_1'^2}{\fg_1^2}+\frac{\fg_2'^2}{\fg_2^2}\Big)
  - g \fg_3 \, \frac{\fg_1' \fg_2' }{\fg_1 \fg_2}, \nn \\
 \mbox{\bf 44:}&\,&  - \,  g^2   \phi'^2+2 \fB^2 g   V+ \frac{ g f_0 A_t'^2}{\fB^2}  + \frac{g f_1 q_1^2 }{\fB^2 \fg_2 \fg_3}
+ \cfrac{g f_2 q_2^2}{\fB^2 \fg_1 \fg_3}
    + \cfrac{g f_3 q_3^2}{\fB^2 \fg_1 \fg_2} \label{44-math} \\  \nn\\
    &&=  - \, \frac{ 24 
    \fB'^2 g^2}{\fB^2} - \cfrac{6 \fB' g' g}{\fB} -\frac{6 \fB' g^2}{\fB} \Big( \frac{\fg_1'}{\fg_1} + 
    \frac{\fg_2'}{\fg_2}+\frac{\fg_3'}{\fg_3}\Big)
    - g' g \Big(\frac{\fg_1'}{\fg_1}
   +\frac{ \fg_2' }{\fg_2} +
   \frac{ \fg_3' }{\fg_3}\Big) \nn \\\nn\\
    &&\quad - \, g^2\Big(\frac{\fg_1' \fg_2' }{\fg_1 \fg_2}+\frac{\fg_1' \fg_3' }{\fg_1 \fg_3}+\frac{\fg_2' \fg_3' }{\fg_2 \fg_3}\Big). \nn
\eea

\section{Ricci and Einstein Tensors in  Parametrization \eqref{eq:2.10}}\label{appC}

Non-zero Christoffel components in parametrization \eqref{eq:2.10}
\begin{gather}
  \Gamma^{\mu}_{4\mu} = \cfrac{g_{00}'}{2 g_{00}}, \ 
  \mu = 0, 1, 2, 3, 4, \quad
  \Gamma^4_{00} = \cfrac{g_{00}'}{2 g_{44}}, \quad
  \Gamma^4_{ii} = - \, \cfrac{g_{ii}'}{2 g_{44}}, \ i = 1, 2, 3. \label{christofg} 
\end{gather}
Ricci tensor and Ricci scalar have the form:
\begin{gather}
  \begin{split}
    R_{00} &= \cfrac{1}{2} \left[
      \cfrac{g_{00}''}{g_{44}}
      - \cfrac{g_{00}'}{g_{44}} \left(
        \cfrac{g_{00}'}{g_{00}} 
        - \sum_{i=1}^3 \cfrac{g_{ii}'}{g_{ii}}
        + \cfrac{g_{44}'}{g_{44}} \right)
    \right], \\
    R_{ii} &= - \, \cfrac{1}{2} \left[
      \cfrac{g_{ii}''}{g_{44}}
      + \cfrac{g_{ii}'}{2 g_{44}} \left(
      \cfrac{g_{00}'}{g_{00}} 
      - \cfrac{g_{ii}'}{g_{ii}}
      + \sum_{j\ne i} \cfrac{g_{jj}'}{g_{jj}}
      - \cfrac{g_{44}'}{g_{44}} \right)
    \right], \\
    R_{44} &= \cfrac{1}{2} \left[
      \cfrac{g_{00}''}{g_{00}}
      + \sum_{i=1}^3 \cfrac{g_{ii}''}{g_{ii}}
      - \cfrac{1}{2} \left( \cfrac{g_{00}'^2}{g_{00}^2}
        + \sum_{i=1}^3 \cfrac{g_{ii}'^2}{g_{ii}^2} \right)
      - \cfrac{g_{44}'}{2 g_{44}} \left( 
        \cfrac{g_{00}'}{g_{00}}
        + \sum_{i=1}^3 \cfrac{g_{ii}'}{g_{ii}} \right)
    \right], \\
    R &= - \, \cfrac{1}{g_{44}} \left[ 
      \cfrac{g_{00}''}{g_{00}}
      + \sum_{i=1}^3 \cfrac{g_{ii}''}{g_{ii}}
      - \cfrac{1}{2} \left( \cfrac{g_{00}'^2}{g_{00}^2}
        + \sum_{i=1}^3 \cfrac{g_{ii}'^2}{g_{ii}^2} \right)
      + \cfrac{g_{00}'}{2 g_{00}} \sum_{i=1}^3 \cfrac{g_{ii}'}{g_{ii}} \right. - \\
      &\quad - \left. 
      \cfrac{g_{44}'}{2 g_{44}} \left( 
        \cfrac{g_{00}'}{g_{00}} 
        + \sum_{i=1}^3 \cfrac{g_{ii}'}{g_{ii}} \right)
      + \cfrac{1}{2} \, \sum_{i=1}^3 \prod_{j\ne i} \cfrac{g_{jj}'}{g_{jj}}
    \right], \ i, j = 1, 2, 3.
  \end{split} \label{ricci}
\end{gather}
The Einstein tensor is
\begin{gather}
    G_{00} = - \cfrac{g_{00}}{2 g_{44}} \left[
    \sum_{i=1}^3 \left( \cfrac{g_{ii}''}{g_{ii}} 
    - \cfrac{g_{ii}'^2}{2 g_{ii}^2} \right)
    - \cfrac{g_{44}'}{2 g_{44}} \sum_{i=1}^3 \cfrac{g_{ii}'}{g_{ii}}
    + \cfrac{1}{2} \sum_{i=1}^3 \prod_{j\ne i} \cfrac{g_{jj}'}{g_{jj}} \right], \ i, j = 1, 2, 3, \label{ein00} \\
    G_{ii} = \cfrac{g_{ii}}{2 g_{44}} \left[
    \cfrac{g_{00}''}{g_{00}} 
    - \cfrac{g_{00}'^2}{2 g_{00}^2} 
    + \sum_{j\ne i} \left( \cfrac{g_{jj}''}{g_{jj}} 
    - \cfrac{g_{jj}'^2}{2 g_{jj}^2} \right)
    - \cfrac{g_{44}'}{2 g_{44}} \left( \cfrac{g_{00}''}{g_{00}} + \sum_{j\ne i} \cfrac{g_{jj}'}{g_{jj}} \right)
    + \cfrac{g_{00}'}{2 g_{00}} \sum_{j\ne i} \cfrac{g_{jj}'}{g_{jj}}
    + \cfrac{1}{2} \prod_{j\ne i} \cfrac{g_{jj}'}{g_{jj}} \right], \label{einii} \\
    G_{44} = \cfrac{1}{2 g_{44}^2} \left[
    \cfrac{g_{00}'}{2 g_{00}} \sum_{i=1}^3 \cfrac{g_{ii}'}{g_{ii}}
    + \cfrac{1}{2} \sum_{i=1}^3 \prod_{j\ne i} \cfrac{g_{jj}'}{g_{jj}} \right]. \label{ein44}
\end{gather}

\section{Stress-Energy Tensor in Parametrization \eqref{eq:2.10}}
\label{appD}

\subsection{Contribution from the Dilaton}\label{appendixD1}

\be
  T_{00}^{\phi} 
  = \cfrac{g_{00}}{2} \left( \cfrac{\phi'^{\, 2}}{2 g_{44}} + V \right), \quad
  T_{ii}^{\phi}
  = - \, \cfrac{g_{ii}}{2} \left( \cfrac{\phi'^{\, 2}}{2 g_{44}} + V \right),
  \ i = 1,2,3, \quad
  T_{44}^{\phi}
  = \cfrac{g_{44}}{2} \left( \cfrac{\phi'^{\, 2}}{2 g_{44}} - V \right). 
  \label{Tphi}
\ee

\subsection{Contribution from the Vector Potential}\label{appendixD2}

\be
  T_{00}^{F_0} 
  = \cfrac{f_0 A_t'^2}{4 g_{44}}, \quad
  T_{ii}^{F_0} 
  = \cfrac{g_{ii} f_0 A_t'^2}{4 g_{00} g_{44}}, \ i = 1,2,3, \quad
  T_{44}^{F_0}
  = - \, \cfrac{f_0 A_t'^2}{4 g_{00}}. \label{TF0}
\ee

\subsection{Full Stress-Energy Tensor}\label{appendixD3}

\bea
    T_{00} &=& \cfrac{g_{00}}{2} \left( 
    \cfrac{\phi'^2}{2 g_{44}} + V 
    + \cfrac{f_0 A_t'^2}{2 g_{00} g_{44}}
    + \cfrac{1}{2} \sum_{i=1}^3 \cfrac{f_i q_i^2}{\prod_{j\ne i} \fg_j}\right), \label{T00} \\
    T_{ii} &=& \cfrac{g_{ii}}{2} \left( 
    - \, \cfrac{\phi'^2}{2 g_{44}} - V 
    + \cfrac{f_0 A_t'^2}{2 g_{00} g_{44}}
    + \cfrac{1}{2} \sum_{j=1}^3 \cfrac{(-1)^{\delta_{ij}} f_j q_j^2}{\prod_{k\ne j} \fg_k} \right), \quad i  = 1, 2, 3, \label{Tii} \\
    T_{44} &=& - \, \cfrac{g_{44}}{2} \left( 
    - \, \cfrac{\phi'^2}{2 g_{44}} + V 
    + \cfrac{f_0 A_t'^2}{2 g_{00} g_{44}}
    + \cfrac{1}{2} \sum_{i=1}^3 \cfrac{f_i q_i^2}{\prod_{j\ne i} \fg_j} \right). \label{T44} 
\eea

\section{EOM in Parametrization \eqref{eq:2.10}}
\label{appE}

\subsection{EOM for Matter Fields}

The EOM for the vector potential has the form:
\begin{gather}
    \cfrac{2 f_0 A_t''}{g_{00} g_{44}}
    + \cfrac{2 A_t' \phi' \partial_{\phi} f_0}{g_{00} g_{44}}
    - \cfrac{g_{00}' f_0 A_t'}{g_{00}^2 g_{44}}
    + \cfrac{g_{11}' f_0 A_t'}{g_{00} g_{11} g_{44}}
    + \cfrac{g_{22}' f_0 A_t'}{g_{00} g_{22} g_{44}}
    + \cfrac{g_{33}' f_0 A_t'}{g_{00} g_{33} g_{44}}
    - \cfrac{g_{44}' f_0 A_t'}{g_{00} g_{44}^2} = 0. \label{eq:At}
\end{gather}
The dilaton EOM has the form:
\begin{gather}
    \begin{split}
    \cfrac{\partial V}{\partial \phi}
    &- \cfrac{\phi''}{g_{44}}
    - \cfrac{g_{00}' \phi'}{2 g_{00} g_{44}}
    - \cfrac{g_{11}' \phi'}{2 g_{11} g_{44}}
    - \cfrac{g_{22}' \phi'}{2 g_{22} g_{44}}
    - \cfrac{g_{33}' \phi'}{2 g_{33} g_{44}}
    + \cfrac{g_{44}' \phi'}{2 g_{44}^2} \, - \\
    &- \cfrac{A_t'^2}{2 g_{00} g_{44}} \, \cfrac{\partial f_0}{\partial \phi}
    + \cfrac{q_1^2}{2 g_{22} g_{33}} \, \cfrac{\partial f_1}{\partial \phi}
    + \cfrac{q_2^2}{2 g_{11} g_{33}} \, \cfrac{\partial f_2}{\partial \phi}
    + \cfrac{q_3^2}{2 g_{11} g_{22}} \, \cfrac{\partial f_3}{\partial \phi} = 0.
    \end{split}\label{eq:phi}
\end{gather}

\subsection{Einstein EOM}

Varying the action \eqref{eq:2.2} with respect to the metric, we obtain the five EOM, which in the parametrization \eqref{eq:2.12} take the form
\bea
    {\bf 00}: 
    &&\sum_{i=1}^3 \left( \cfrac{g_{ii}''}{g_{ii}}
      - \cfrac{g_{ii}'^2}{2g_{ii}^2} \right)
      - \cfrac{1}{2} \left(
        \cfrac{g_{44}'}{g_{44}} \sum_{i=1}^3 \cfrac{g_{ii}'}{g_{ii}}
        - \sum_{i=1}^3 \prod_{j=1,j\ne i}^3 \cfrac{g_{jj}'}{g_{jj}} \
       \right) + \label{eom00} \\
      &&\qquad \quad \ \ + \, \cfrac{g_{44}}{2} \left(
        \cfrac{f_0 A_t'^2}{g_{00} g_{44}}
        + \sum_{i=1}^3 \cfrac{f_i q_i^2}{\prod_{j\ne i} g_{jj}}
        + \cfrac{\phi'^2}{g_{44}^2} + 2 V
      \right) = 0, 
\eea
\bea
    {\bf ii}: &&\cfrac{g_{00}''}{g_{00}} - \cfrac{g_{00}'^2}{2g_{00}^2} + \sum_{j=1,j\ne i}^3 \left( \cfrac{g_{jj}''}{g_{jj}}
      - \cfrac{g_{jj}'^2}{2g_{jj}^2} \right) \nn\\
      &&\quad \ \ - \, \cfrac{1}{2} \left[
        \cfrac{g_{44}'}{g_{44}} \left( \cfrac{g_{00}'}{g_{00}} + \sum_{j=1,j\ne i}^3 \cfrac{g_{jj}'}{g_{jj}} \right)
        - \left(
          \cfrac{g_{00}'}{g_{00}} \, \sum_{j=1,j\ne i}^3 \cfrac{g_{jj}'}{g_{jj}}
          + \prod_{j=1,j\ne i}^3 \cfrac{g_{jj}'}{g_{jj}}
        \right) 
    \right] + \label{eq:3.04} \\
     &&\quad \ \ + \, \cfrac{g_{44}}{2} \left(
        - \, \cfrac{f_0 A_t'^2}{g_{00} g_{44}}
        - \sum_{i=1}^3 (- 1)^{\delta_{ij}} f_j q_j^2\prod_{k=1,k\ne j}^3\frac{1}{g_{kk}}
        + \cfrac{\phi'^2}{g_{44}^2} + 2 V
      \right) = 0, \nn \\
   {\bf 44}: &&\cfrac{g_{00}'}{2g_{00}}
      \sum_{i=1}^3 \cfrac{g_{ii}'}{g_{ii}}
      + \cfrac{1}{2} \sum_{i=1}^3 \prod_{j=1,j\ne i}^3 \cfrac{g_{jj}'}{g_{jj}}
      + \cfrac{g_{44}}{2} \left(
        \cfrac{f_0 A_t'^2}{g_{00} g_{44}}
        + \sum_{i=1}^3 f_i q_i^2\prod_{j=1,j\ne i}^3\frac{1} {g_{jj}}
        - \cfrac{\phi'^2}{g_{44}^2} + 2 V
      \right) = 0. \nn \\
      && \ \label{eom44}
\eea

\subsection{Useful Form of Einstein EOM}\label{app:UFEOM}

We can see that equations \eqref{eom00}--\eqref{eom44} have a rather complicated form on the one hand and include repeating combinations of terms on the other hand. For further analysis, let us combine these Einstein equations into linear combinations, thus excluding the repeating terms and focusing on the specific details. To achieve this, we use the procedure that consists of writing the following combination of EOM (where bold double indices label the EOM as indicated in \eqref{eom00}--\eqref{eom44}):
\bea
&&   \cfrac{2 g_{44}}{\sqrt{- g}} \left(
      - \, \cfrac{\bf 00}{g_{00}} + \cfrac{\bf 11}{g_{11}} \right),\quad \cfrac{2 g_{44}}{\sqrt{- g}} \left(
      \cfrac{\bf 00}{g_{00}} - \cfrac{\bf 44}{g_{44}} \right),\quad\cfrac{- 2 g_{44}}{\sqrt{- g}} \left(
      \cfrac{\bf 11}{g_{11}} - \cfrac{\bf 22}{g_{22}} \right),\nn \\
    &&\cfrac{- 2 g_{44}}{\sqrt{- g}} \left(
      \cfrac{\bf 11}{g_{11}} - \cfrac{\bf 33}{g_{33}} \right),\quad 
    \cfrac{2 g_{44}}{\sqrt{- g}} \left(
      \cfrac{\bf 22}{g_{22}} + \cfrac{\bf 44}{g_{44}} \right).
 \label{linear}
\eea

This procedure yields five equations in the form
\begin{gather}
  \begin{split}
  \cfrac{g_{00}''}{g_{00}}
  &- \cfrac{g_{00}'^2}{2 g_{00}^2}
  - \left(\cfrac{g_{11}''}{g_{11}}
  - \cfrac{g_{11}'^2}{2 g_{11}^2} \right)
  + \cfrac{1}{2} \left(
    \cfrac{g_{00}'}{g_{00}} - \cfrac{g_{11}'}{g_{11}} 
  \right)
  \left(
    \cfrac{g_{22}'}{g_{22}} + \cfrac{g_{33}'}{g_{33}} 
    - \cfrac{g_{44}'}{g_{44}}
  \right) - \\
  &- \cfrac{f_0 (A_t')^2}{g_{00}}
  - \cfrac{g_{44} \, f_2 q_2^2}{g_{11} g_{33}}
  - \cfrac{g_{44} \, f_3 q_3^2}{g_{11} g_{22}} = 0,
  \end{split}\label{linear1} \\
  \sum_{i=1}^3 \left( \cfrac{g_{ii}''}{g_{ii}}
  - \cfrac{g_{ii}'^2}{2g_{ii}^2}
  \right)
  - \left( 
      \cfrac{g_{00}'}{g_{00}} + \cfrac{g_{44}'}{g_{44}}
    \right)
    \sum_{i=1}^3 \cfrac{g_{ii}'}{2g_{ii}}
   + \phi'^2 = 0, \quad i = 1, 2, 3, \label{linear2}
\end{gather}
\begin{gather}
  \cfrac{g_{11}''}{g_{11}}
  - \cfrac{g_{11}'^2}{2 g_{11}^2}
  - \left( \cfrac{g_{22}''}{g_{22}}
  - \cfrac{g_{22}'^2}{2 g_{22}^2} \right)
  + \cfrac{1}{2} \left(
    \cfrac{g_{11}'}{g_{11}} - \cfrac{g_{22}'}{g_{22}} 
  \right)
  \left(
    \cfrac{g_{00}'}{g_{00}} + \cfrac{g_{33}'}{g_{33}} 
    - \cfrac{g_{44}'}{g_{44}}
  \right)
  - \cfrac{g_{44} \, f_1 q_1^2}{g_{22} g_{33}}
  + \cfrac{g_{44} \, f_2 q_2^2}{g_{11} g_{33}} = 0,
  \label{linear3} \\
  \cfrac{g_{11}''}{g_{11}}
  - \cfrac{g_{11}'^2}{2 g_{11}^2}
  - \left( \cfrac{g_{33}''}{g_{33}}
    - \cfrac{g_{33}'^2}{2 g_{33}^2} \right)
  + \cfrac{1}{2} \left(
    \cfrac{g_{11}'}{g_{11}} - \cfrac{g_{33}'}{g_{33}} 
  \right)
  \left(
    \cfrac{g_{00}'}{g_{00}} + \cfrac{g_{22}'}{g_{22}} 
    - \cfrac{g_{44}'}{g_{44}}
  \right)
  - \cfrac{g_{44} \, f_1 q_1^2}{g_{22} g_{33}}
  + \cfrac{g_{44} \, f_3 q_3^2}{g_{11} g_{22}} = 0,
  \label{linear4} \\
  \begin{split}
    \cfrac{g_{00}''}{g_{00}} + \cfrac{g_{11}''}{g_{11}} 
    + \cfrac{g_{33}''}{g_{33}}
    &+ \cfrac{g_{00}'}{2 g_{00}} \left(
      - \, \cfrac{g_{00}'}{g_{00}} + \sum_{i=1}^3 \cfrac{g_{ii}'}{g_{ii}}
      - \cfrac{g_{44}'}{g_{44}}
    \right)
    + \cfrac{g_{11}'}{2 g_{11}} \left(
      \cfrac{g_{00}'}{g_{00}} - \cfrac{g_{11}'}{g_{11}}
      + \cfrac{g_{22}'}{g_{22}} + \cfrac{g_{33}'}{g_{33}}
      - \cfrac{g_{44}'}{g_{44}}
    \right) + \\
    &+ \cfrac{g_{33}'}{2 g_{33}} \left(
      \cfrac{g_{00}'}{g_{00}} + \cfrac{g_{11}'}{g_{11}}
      + \cfrac{g_{22}'}{g_{22}} - \cfrac{g_{33}'}{g_{33}}
      - \cfrac{g_{44}'}{g_{44}}
    \right)
    + \cfrac{g_{44} \, f_2 q_2^2}{g_{11} g_{33}} 
    + 2 g_{44} V = 0.
  \end{split}\label{linear5}
\end{gather}

\end{document}